\documentclass{PoS}
\usepackage{graphics,psfrag}
\usepackage{graphicx,psfrag}
\title{Quark localization in QCD above $T_c$}

\ShortTitle{Quark localization in QCD above $T_c$}
\author{Matteo Giordano\thanks{Supported by the Hungarian Academy of Sciences under ``Lend\"ulet'' grant No. LP2011-011. TGK and FP acknowledge partial support by the EU Grant (FP7/2007
-2013)/ERC No. 208740. We also thank the Budapest-Wuppertal group for allowing us to use their code to generate
the gauge configurations.}\\
  Institute for Nuclear Research of the Hungarian Academy of
  Sciences,\\
  Bem t\'er 18/c, H-4026 Debrecen, Hungary\\
  E-mail: \email{giordano@atomki.mta.hu}}
\author{Tam\'as G. Kov\'acs\footnotemark[1]\\
  Institute for Nuclear Research of the Hungarian Academy of
  Sciences,\\
  Bem t\'er 18/c, H-4026 Debrecen, Hungary\\
  E-mail: \email{kgt@atomki.mta.hu}}
\author{\speaker{Ferenc Pittler}\footnotemark[1]\\
  Institute for Nuclear Research of the Hungarian Academy of
  Sciences,\\
  Bem t\'er 18/c, H-4026 Debrecen, Hungary\\
  E-mail: \email{pittler@atomki.mta.hu}}

\abstract{It was previously found that at high temperature the lowest part of the 
QCD Dirac spectrum consists of localized modes obeying Poisson statistics. Higher 
up in the spectrum, modes become delocalized and their statistics can be described 
by random matrix theory. The transition from localized to delocalized modes is analogous 
to the Anderson metal-insulator transition. Here we use dynamical QCD simulations with 
staggered quarks to study this localization phenomenon. We show that the ``mobility edge'', 
separating localized and delocalized modes, scales properly in the continuum limit and 
rises steeply with the temperature. Using very high statistics simulations in large volumes 
we find that the density of localized modes scales precisely with the spatial volume and even 
at $T=2.6T_{c}$ the lowest part of the spectrum extends all the way down to zero with no evidence 
of a spectral gap.}

\FullConference{31st International Symposium on Lattice Field Theory - LATTICE 2013\\
		July 29 - August 3, 2013\\
		Mainz, Germany}

\begin{document}

\section{Introduction}

Although not being directly accessible to experiments, the spectrum of
the Dirac operator, and in particular its low end, contains important
information on observable properties of QCD. For example, the order
parameter of the chiral phase transition, the chiral condensate, is
related to the density of low eigenvalues of the Dirac operator
through the Banks-Casher relation~\cite{Banks:1979yr}. Moreover, as
the quark propagator is the inverse of the Dirac operator, the lowest
Dirac modes get the largest weight in its mode decomposition.


In QCD at low temperatures, the low-lying eigenmodes of the Dirac 
operator are delocalized, and in the so-called \textit{epsilon-regime}
the corresponding eigenvalues are well described by the Wigner-Dyson
statistics of Random Matrix Theory
(RMT)~\cite{Shuryak:1992pi,BerbenniBitsch:1997tx}. However, this is no
longer true at high temperature. It was shown
in~\cite{GarciaGarcia:2005vj} that around the critical temperature
$T_c$ the eigenmodes  of the staggered Dirac operator become localized near the ``spectrum edge'',
i.e., near $\lambda=0$. 
Moreover, the first few eigenvalues of the overlap Dirac operator in pure $SU(2)$  gauge 
theory were found to be statistically independent, following Poisson statistics \cite{Kovacs:2009zj}. 
In a previous paper, we showed that both Poisson and Wigner-Dyson statistics appear in the 
staggered Dirac spectra in pure $SU(2)$ gauge theory, in different spectral windows separated by 
the so-called ``mobility edge'' \cite{Kovacs:2010wx}. More recently, we showed that this
transition appears also in the case of real QCD with 2+1 flavors of dynamical quarks with 
physical quark masses \cite{Kovacs:2012zq}. In this paper we make large improvement on the 
statistics compared to \cite{Kovacs:2012zq}. We also show that while it is unlikely that a
true spectral gap is present in the Dirac spectrum at high temperature, nevertheless an 
effective gap appears due the presence of the low-lying localized modes.


\section{Simulation details}

We diagonalize the staggered Dirac operator in SU(3) gauge theory with 2+1 flavors of dynamical 
quarks. We use physical quark masses in our simulations
\cite{Aoki:2006br}. To determine precisely the spectral density at the
low end of the spectrum we generated large ensembles on large lattice
volumes (see Table~\ref{tab:extension}). We have ensembles in the
temperature range $\left[260\mathrm{~MeV},800\mathrm{~MeV}\right]$. We
use three different lattice spacings $a=0.125$ fm, $0.082$ fm, $0.062$
fm in order to estimate the scaling violations. 


\begin{table}
\centering
\begin{tabular}{c | c | c}
$N_{s}$ & $N_{conf}$ & $N_{eig}$ \\
\hline
24 & 45122 & 256 \\
36 & 14445 & 550 \\
48 & 6984  & 1000 \\
\end{tabular}
\caption{\label{tab:extension}Details of our ensembles at lattice 
spacing $a=0.125$ fm, and temperature $T=394$ MeV. 
$N_{s}$ is the linear spatial extension of the lattice, 
$N_{conf}$ is the number of generated configurations and 
$N_{eigv}$ is the number of eigenvalues determined for each 
configuration.}
\end{table}

\section{Results}

\begin{figure}
\psfrag{yaxis}[b][b][0.7]{$\langle PR\rangle=\langle\frac{1}{V}\left(
\sum_{x}\vert\psi_{i}^\dagger\left(x\right) \psi_{i}\left(x\right) \vert^{2}\right)
^{-1}\rangle$}
\psfrag{xaxis}[b][b][0.7]{$\lambda a$}
\psfrag{24}[b][b][0.9]{$~~~~~~N_{s}=24$}
\psfrag{36}[b][b][0.9]{$~~~~~~N_{s}=36$}
\psfrag{48}[b][b][0.9]{$~~~~~~N_{s}=48$}
\centering
\includegraphics[width=0.48\textwidth,clip=true]{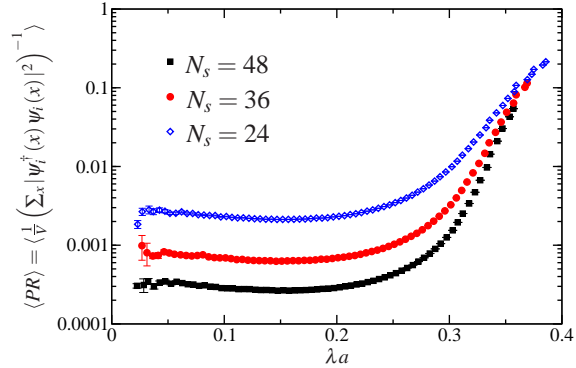}
\caption{The local average participation ratio of eigenvectors for three different volumes. Here $T= 394 MeV$.}
\label{participation_ratio}
\end{figure}

We first show that in QCD at high temperature the eigenmodes of the
Dirac operator behave quite differently depending on whether they are  
located in the bulk or at the edge of the spectrum. The localization
properties of an eigenmode $\psi_{i}$ can be examined by studying the
so-called participation ratio $PR=\frac{1}{V}\left(\sum_{x}\vert
  \psi_{i}^\dagger\left(x\right)\psi_{i}\left(x\right)\vert^{2}\right)^{-1}$,
where $V$ is the spatial volume. The $PR$ essentially measures the
fraction of the lattice volume occupied by a given eigenmode. In the
thermodynamic limit, the average $PR$ for localized modes is zero,
while for delocalized modes it is a non-zero finite number. In
Fig. \ref{participation_ratio} we show how the average $PR$ changes
along the spectrum 
for three different lattice volumes. 
The low-lying modes occupy only a small fraction of the total volume, 
which decreases when increasing the linear spatial size of the lattice
$N_{s}$, meaning that they are localized. In contrast, the eigenmodes
in the bulk occupy the same fraction of the total volume independently
of the volume, i.e., they are delocalized. 


\begin{figure}
\begin{center}

\begin{tabular}{cc}
\includegraphics[width=0.48\textwidth,clip=true]{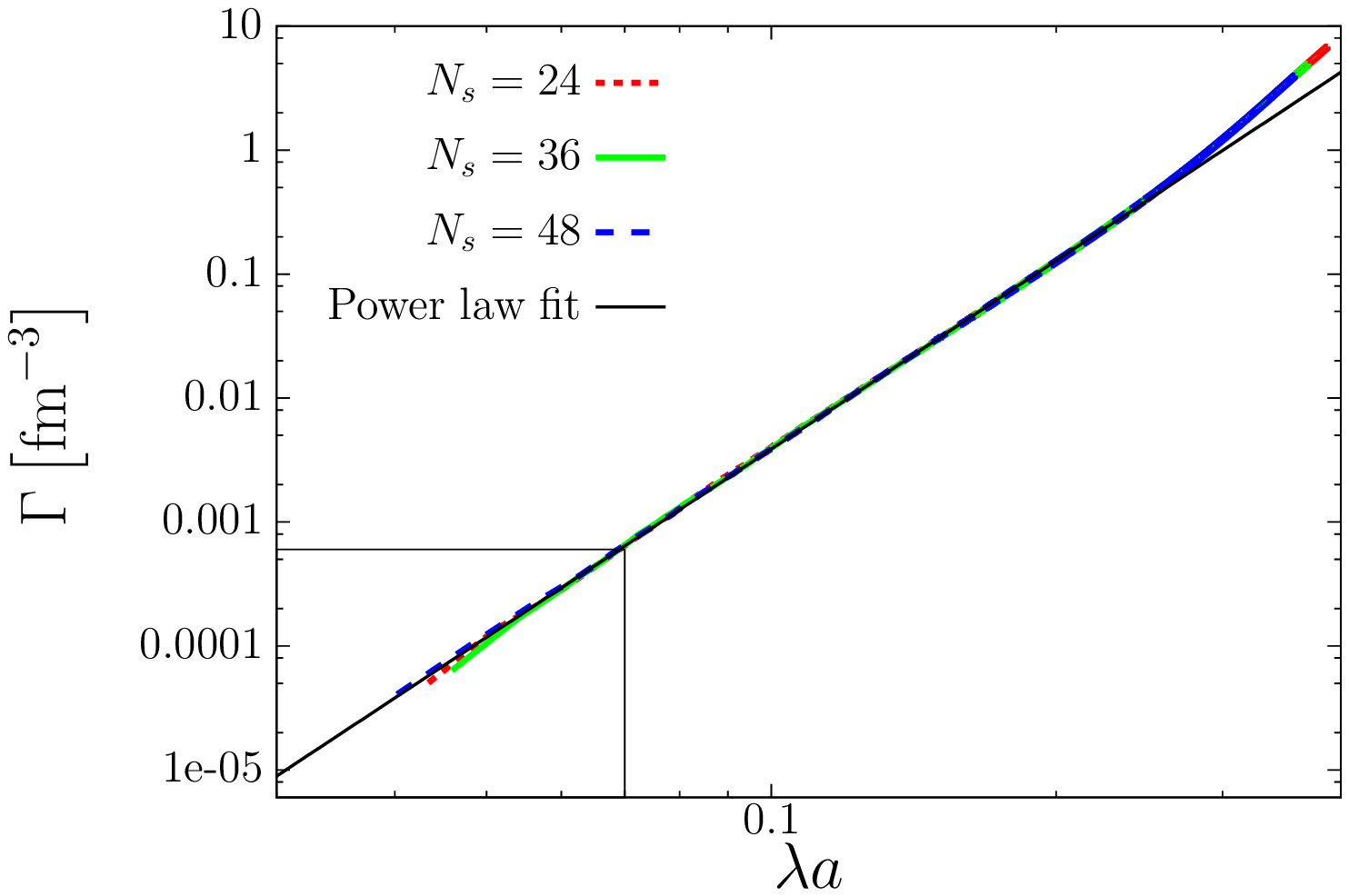}
& 
\includegraphics[width=0.48\textwidth,clip=true]{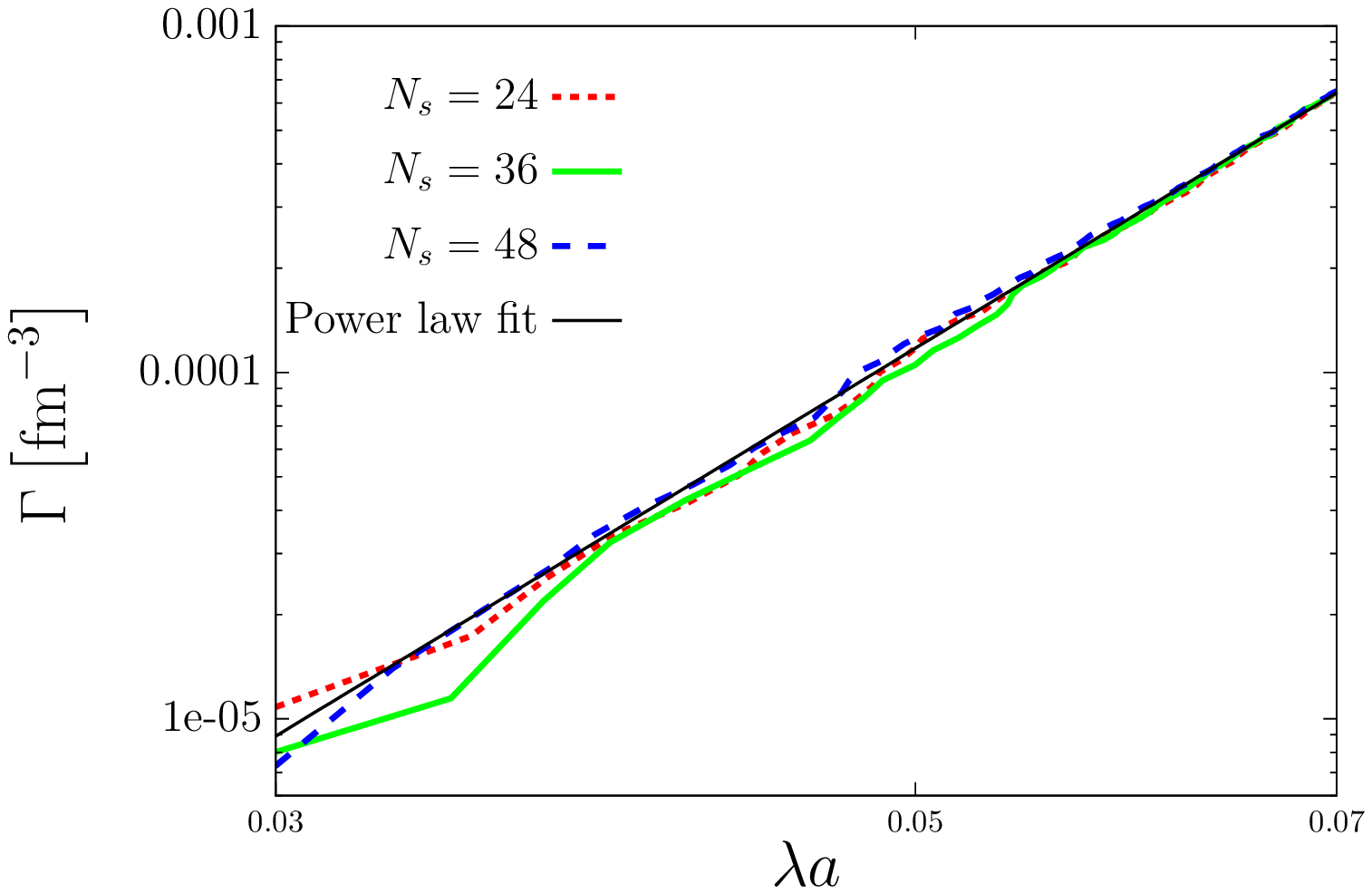}
\end{tabular}
\end{center}
\caption{Left panel: The integrated spectral density normalized by the
  volume
  for the ensembles listed in Tab.~\protect\ref{tab:extension}. The 
  continuous line corresponds to a power law fit to the data on 
     the largest volume ($N_{s}=48$) in the interval 
     $\left[0.09,0.12\right]$. Right panel: Zoom in the spectral 
     region around $\lambda\simeq 0$ i.e. around
     the ``spectrum edge''. }
\label{spectral_density}
\end{figure}

Localized modes appear at the low end of the spectrum, where the
spectral density is small. In Fig.~\ref{spectral_density} we show our
results for the integrated spectral density of the Dirac
operator normalized by the spatial volume, $\Gamma=\frac{1}{V}\int_{0}^{\lambda a} \rho\left(\lambda'
  a\right)\mathrm{d}(\lambda' a)$. As
curves corresponding to different volumes lie on top of each other, we
conclude that the spectral density scales with $V$. 
While it is clear that $\Gamma$ vanishes rapidly as one
approaches the origin, there are no indications of non-analyticity. 
Indeed, a power-law fit to the data in the interval
$\left[0.09,0.12\right]$ works very well also below the fitting range.
We then expect that the spectral density behaves smoothly in the limit
$\lambda\to 0$. From the fit we get that the spectral density vanishes 
as $\lambda^p$ with $p=4.047\pm0.001$ at the spectrum edge, which is
faster than in the free case ($\rho\left(\lambda\right)\sim
\lambda^{3}$). 

The combination of localization and small spectral density leads one to
expect that the low-lying eigenmodes are not mixed by the fluctuations
of the gauge field. As a consequence, the corresponding eigenvalues
are expected to be uncorrelated, thus following Poisson statistics.
Notice that low spectral density alone does not necessarily lead to
uncorrelated eigenvalues: this is the case in the two-flavor Schwinger
model \cite{Bietenholz:2013coa}. 
Moving towards the bulk of the spectrum, the density of eigenvalues
increases rapidly, and the eigenvectors start to become 
delocalized. The typical gauge field fluctuations are then expected to
easily mix the eigenmodes, as their energy difference is small
and they also have significant spatial overlap. In this case, the
corresponding eigenvalues are expected to follow the Wigner-Dyson
statistics of the appropriate ensemble of RMT, which is the unitary
ensemble in the case of QCD.

\begin{figure}
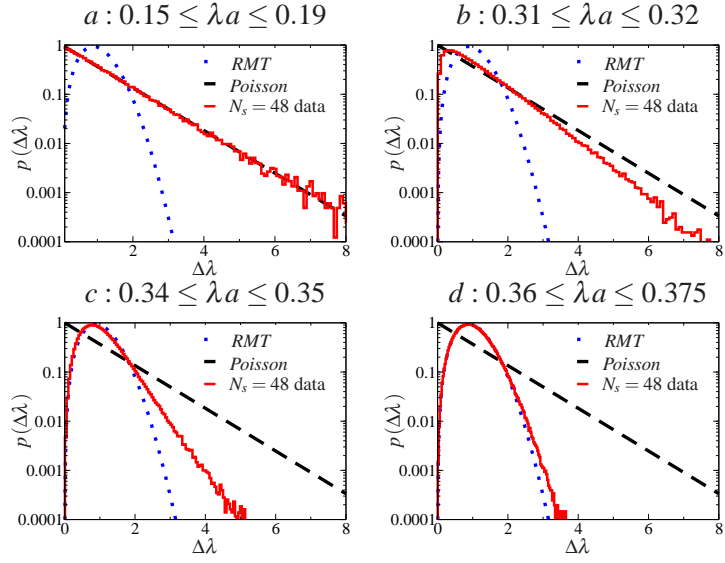

\centering
\psfrag{RMT}[b][b][0.65]{$~~RMT$}
\psfrag{Poisson}[b][b][0.65]{$Poisson$}
\psfrag{48}[b][b][0.65]{$~~~~~~~~~~~~~~~~~~~N_{s}=48$ data}
\psfrag{a}[b][b][1]{$a:0.15\le \lambda a \le 0.19$}
\psfrag{b}[b][b][1]{$b:0.31\le \lambda a \le 0.32$}
\psfrag{c}[b][b][1]{$c:0.34\le \lambda a \le 0.35$}
\psfrag{d}[b][b][1]{$d:0.36\le \lambda a \le 0.375$}
\psfrag{yaxis}[b][b][0.7]{$p\left(\Delta \lambda\right)$}
\psfrag{xaxis}[b][b][0.7]{$\Delta \lambda$}
\begin{tabular}{cc}
\includegraphics[width=0.3\columnwidth,keepaspectratio]{Figs/0_15_0_19_484_pos.eps}&

\includegraphics[width=0.3\columnwidth,keepaspectratio]{Figs/0_31_0_32_484_pos.eps}\\

\includegraphics[width=0.3\columnwidth,keepaspectratio]{Figs/0_34_0_35_484_pos.eps}&

\includegraphics[width=0.3\columnwidth,keepaspectratio]{Figs/0_36_03735_484_pos.eps}
\\
\end{tabular}
\caption{\label{fig:unfold}The unfolded level spacing ($\Delta\lambda = 
\frac{\lambda_{n+1}-\lambda_{n}}{\langle \lambda_{n+1}-\lambda_{n} \rangle}$) 
distribution (ULSD) in four non-overlapping spectral windows. The
dashed and dotted line indicates the ULSD for Poisson statistics 
and for RMT statistics respectively. The spatial extension of the 
lattice is $N_{s}=48$ and the temperature is $T=394$~MeV.}
\end{figure}

To detect the expected transition in the spectral statistics
we use the distribution of spacings between neighboring eigenvalues on the scale 
of the local average level spacing, the so-called unfolded level spacing distribution (ULSD). 
For Poisson statistics 
the ULSD is a simple exponential, while for 
Wigner-Dyson statistics it is well approximated by the so-called Wigner surmise
for the appropriate random matrix ensemble. We display our results in
Fig.~\ref{fig:unfold}. For comparison we include the exponential
(dotted line)  and the unitary Wigner surmise (continuous line) in
each panel. In panel $a$\footnote{The reason  
why we do not show our results below this interval in the spectrum is simply that
we do not have enough eigenvalues in that spectral window to obtain
a good statistics (see Fig.~\ref{spectral_density}).} we see a nice agreement with the
exponential, indicating that the low modes follow Poisson statistics. In
panel $b$ we show the ULSD in a spectral window where the spectral
density is one order of magnitude larger. There is a clear signal of a
deviation from the Poisson exponential towards the unitary Wigner
surmise, which increases as one moves further up along the spectrum
(panel $c$). Approaching even more the bulk of the spectrum (panel
$d$) the spectral statistics nearly agrees with the Wigner surmise.  
Thus the eigenvalue statistics in the bulk of the Dirac spectrum is
well described by Random Matrix Theory in the high
temperature quark-gluon-plasma phase as well as in the low temperature
hadronic phase.


\begin{figure}
\centering
\psfrag{yaxis}[b][b][0.8]{$\sigma^{2}\left(\lambda a\right)$}
\psfrag{xaxis}[b][b][0.8]{$\lambda a$}
\psfrag{24}[b][b][0.8]{$~~~~~~N_{s}=24$}
\psfrag{36}[b][b][0.8]{$~~~~~~N_{s}=36$}
\psfrag{48}[b][b][0.8]{$~~~~~~N_{s}=48$}
\psfrag{Poisson}[b][b][0.8]{Poisson}
\psfrag{RMT}[b][b][0.8]{~RMT}
\includegraphics[width=0.47\textwidth,clip=true]{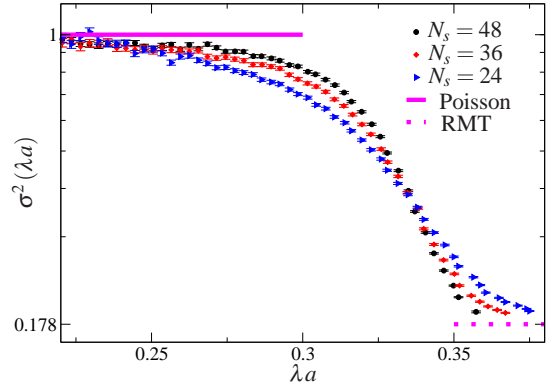}
\caption{The variance of the local ULSD for three different volumes. The smooth and 
the dashed line indicates the variance of ULSD in the case of Poisson statistics and 
RMT statistics respectively.}
\label{ulsd_vol_dep}
\end{figure}

It is instructive to check the volume dependence of the transition in
the spectrum. Instead of comparing the whole distribution in a  given
spectral window on different volumes, it is simpler to just pick one
parameter of the distribution and see how it changes along 
the spectrum. For this purpose we use the variance of the ULSD, which
can be computed analytically for both kinds of statistics.\footnote{It
is 1 for Poisson statistics and approximately 0.178 for the unitary 
ensemble of RMT.}  We show our results in Fig.~\ref{ulsd_vol_dep}. It is clearly
seen that the transition becomes sharper for larger volumes, which
suggests that it becomes a true phase transition in the thermodynamic
limit. A proper finite size scaling analysis of the transition and a
comparison with the Anderson transition is discussed in
\cite{Matteos-talk}. In a finite volume the separation between localized
and delocalized modes is not sharp, and so the definition of the
``mobility edge'' $\lambda_c$ is not unique. Here we define it to be 
the inflection point of 
the variance of the ULSD. If there is a genuine phase transition, 
this point should tend to the true critical point in the thermodynamic limit.
The critical statistics is examined in~\cite{Nishigaki}.  

As long as hadronic correlators are concerned, the ``mobility edge''
$\lambda_c$ acts as an effective gap in the spectrum. To see this
explicitly, let us write the quark propagator in terms of the Dirac
eigenmodes 
\begin{equation}
\label{quarkprop}
G\left(x,y\right)=\sum_{n}\frac{1}{i\lambda_{n}+m_{q}}\psi_{n}\left(x\right)
\psi^{\dagger}_{n}
\left(y\right),
\end{equation}
where $x,y$ are space time points, $m_{q}$ is the bare quark mass and the sum is 
over all eigenmodes $\psi_{n}$ of the Dirac operator. Firstly, the sum in (\ref{quarkprop}) 
is dominated by the low eigenmodes, which are localized. 
On the other hand, for a localized mode, the product 
$\psi\left(x\right)\psi^{\dagger}\left(y\right)$ is practically zero
when the distance between $x$ and $y$ is larger than the localization
length, 
and so localized modes do not propagate quarks to large
distances. Therefore, long range hadronic correlators
receive contributions only from the delocalized 
eigenmodes, for which $\lambda >\lambda_c$. 

\begin{figure}
\psfrag{yaxis}[b][b][1]{$\langle PR\rangle=\langle\frac{1}{V}\left(\sum_{x}\vert \psi_{i}\left(x\right)
\psi_{i}^\dagger\left(x\right)\vert^{2}\right)
^{-1}\rangle$}
\psfrag{xaxis}[b][b][1]{$a\lambda$}
\psfrag{24}[b][b][1.25]{$~~~~~~N_{s}=24$}
\psfrag{36}[b][b][1.25]{$~~~~~~N_{s}=36$}
\psfrag{48}[b][b][1.25]{$~~~~~~N_{s}=48$}
\psfrag{Poisson}[b][b][1.25]{Poisson}
\psfrag{RMT}[b][b][1.25]{RMT}
\centering
\includegraphics[width=0.47\textwidth,clip=true]{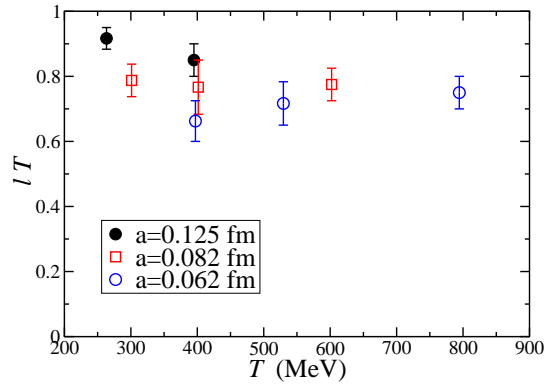}
\caption{The temperature dependence of the localization length in units of 
the inverse temperature for three different lattice spacings.}
\label{loclength}
\end{figure}

A possible definition of the localisation length $l$ of localized modes is just 
$l=a(V\cdot PR)^{\frac{1}{4}}=a\left(\sum_{x}\vert \psi^{\dagger}\left(x\right)\psi
\left(x\right)\vert^{2}\right)^{-\frac{1}{4}}$. In Fig.~\ref{loclength} we show 
$l$ in units of the inverse temperature for three 
different lattice spacings. The localization length in units of the
inverse temperature remains around unity for each lattice
spacing. This suggests that the length scale of the localized modes is
set by the inverse temperature. The low localized eigenmodes are
squeezed in the spatial directions in the same way as in the temporal
direction.  

\begin{figure}
\centering
\includegraphics[width=0.48\textwidth,clip=true]{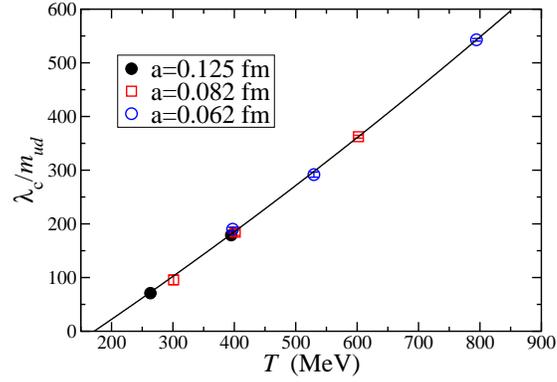}
\caption{The rescaled ``mobility edge'' $\lambda^{rs}_{c}=\frac{\lambda_{c}}{m_{ud}}$ as a function of the temperature for three different 
lattice spacings.}
\label{renormmob}
\end{figure}

The ``mobility edge'' behaves as it were effectively a gap in the spectrum with respect 
to the long range correlators. To determine how this effective gap depends on the temperature 
we define a rescaled ``mobility edge'' $\lambda^{rs}_{c}\equiv\frac{\lambda_{c}}{m_{ud}}$, 
where $m_{ud}$ is the bare quark mass. This quantity has a well defined continuum limit 
\cite{Kovacs:2012zq}. We show our results for the rescaled ``mobility edge'' in 
Fig.~\ref{renormmob} as a function of the temperature for three different lattice spacings. 
The points fall nearly on a straight line. The systematic error arising from finite lattice 
spacing effects are comparable to our statistical errors.  To determine the possible deviations
from the linear behavior of $\lambda^{rs}_{c}$ we make a second order polynomial fit ($\chi^{2}=1.32$). 
The coefficient of the first order term in $\frac{T-T_{c}}{T_{c}}$ is two orders of magnitude larger 
than the coefficient of the second order term, so to a good approximation  $\lambda^{rs}_{c}$ 
increases linearly with the temperature. Notice that the ``mobility edge'' is already 
two orders of magnitude larger than the quark mass just above $T_{c}$. We get a good crosscheck for 
our results by extrapolating this fit to $\lambda^{rs}_{c}$ equal to zero,  which corresponds to 
the temperature at which the localized modes start to appear. We find for this point T$\simeq$170 MeV, which is 
above the point where the finite temperature chiral crossover takes place \cite{Borsanyi:2010bp,Bazavov:2011nk},
consistently with the absence of localised modes at low temperature.


\section{Conclusion}

We have shown that even at temperatures well above $T_c$ 
the Dirac operator seems to have no spectral gap. However, the low-lying modes 
behave quite differently compared to the modes in the bulk: they are
localized on the scale of the inverse temperature, and the
corresponding eigenvalues are statistically independent. Above the
``mobility edge'', the eigenmodes are delocalized, as in the low
temperature regime. This ``mobility edge'' plays the role of an
effective gap with respect to the long range correlators. Here long
range means that the separation of the fields in the correlator is
larger than the inverse temperature.  


\end{document}